# Single-shot readout of the nuclear spin of an on-surface atom

Evert W. Stolte[1,*], Jinwon Lee[1,*,†], Hester G. Vennema[1], Rik Broekhoven[1], Esther Teng[1], Allard J. Katan[1], Lukas M. Veldman[2], Philip Willke[3], Sander Otte[1,‡]

[1]Department of Quantum Nanoscience, Kavli Institute of Nanoscience, Delft University of Technology, Delft, The Netherlands
[2]Institute for Functional Matter and Quantum Technologies, University of Stuttgart, Stuttgart, Germany
[3]Physikalisches Institut, Karlsruhe Institute of Technology, Karlsruhe, Germany
[*] These authors contributed equally
[†] jinwon.lee@tudelft.nl
[‡] a.f.otte@tudelft.nl

**Abstract:**

Nuclear spins owe their long-lived magnetic states to their excellent isolation from the environment. At the same time, a finite degree of interaction with their surroundings is necessary for reading and writing the spin state. Therefore, detailed knowledge of and control over the atomic environment of a nuclear spin is key to optimizing conditions for quantum information applications. While various platforms enabled single-shot readout of nuclear spins, their direct environments were either unknown or impossible to controllably modify on the atomic scale. Scanning tunnelling microscopy (STM), combined with electron spin resonance (ESR), provides atomic-scale information of individual nuclear spins via the hyperfine interaction. Here, we demonstrate single-shot readout of an individual $^{49}$Ti nuclear spin with an STM. Employing a pulsed measurement scheme, we find its lifetime to be in the order of seconds. Furthermore, we shed light on the pumping and relaxation mechanisms of the nuclear spin by investigating its response to both ESR driving and tunnelling current, which is supported by model calculations. These findings give an atomic-scale insight into the nature of nuclear spin relaxation and are relevant for the development of atomically assembled qubit platforms.

**Introduction**

Nuclear spins are unique in that they combine subatomic length scales with generally long coherence times. These properties make them appealing candidates for quantum technological applications, while at the same time posing challenges for their readout. Ideally, one would detect the time-dependent behavior of a spin via a single-shot readout, as opposed to a time-averaged measurement. This enables systematically linking specific changes in an open quantum system to observed or controlled changes in its environment, with the potential for real-time feedback. Single-shot readout of individual nuclear spins was achieved in a number of platforms: optically addressed colour centres[1-3], molecular break junctions[4], dopants in semiconductors[5,6], and gate-defined quantum dots[7]. While it was shown to be possible to trace the spins in these systems over time with excellent accuracy, their direct environments were either unknown or impossible to controllably modify on the atomic scale.

Spins in individual atoms on surfaces, probed through scanning tunneling microscopy (STM), can be positioned with atomic precision[8]. This allows for engineering the environment, including spin-spin[9] as well as spin-orbit interactions[10,11], and for building desired atomic-scale spin structures for the purpose of atomic scale quantum simulation experiments[12-14]. Moreover, combining STM with electron spin resonance (ESR)[15] has enhanced the energy resolution and enabled a wealth of detailed studies on composite spin systems[16-20] including their quantum coherent evolution over time[21-23]. Recently, access to the nuclear spin was provided through resolving the hyperfine interaction[24], potentially introducing spins with much longer coherence times to the quantum simulation efforts using atoms on surfaces. While it has been shown that ESR-STM allows for both reading[25,26] and directly driving nuclear spins[27], STM has thus far only sparingly been used to investigate nuclear spins in the time domain[28]. As such, no nuclear spin lifetimes have yet been reported and it remains unclear what the dominant relaxation processes are.

In this work, we demonstrate single-shot readout of the nuclear spin state of a single Ti atom, expanding on ESR-STM methodology. We apply a fixed radio frequency (RF), tuned to drive the electron spin if and only if the nuclear spin has magnetization $m_I = -7/2$. If this condition is met, spin-polarized transport through the electron orbital results in a measurable increase in the tunnelling current $I_{ESR}$. We observe that the current switches between its base value $I_0$ and the increased value $I_0 + I_{ESR}$ on the timescale of seconds. From this we determine the intrinsic lifetime of the nuclear spin to be 5.3 ± 0.5 seconds, seven orders of magnitude longer than the lifetime of the electron spin on the same atom[16,22]. In addition, we investigate how the lifetime is affected by the presence of tunneling electrons or a continuous RF signal.

## Results and Discussion

### Single-shot readout

We investigate individual Ti atoms on the oxygen binding site of an MgO/Ag(100) substrate (see Methods), which have been shown to carry an electron spin $S = 1/2$ that can be ESR driven by an RF voltage at the STM junction[16], in an external magnetic field oriented out-of-plane. Ti has different isotopes carrying nuclear spins 0, 5/2, or 7/2. In particular, $^{49}$Ti has nuclear spin $I = 7/2$ and couples to the electron spin with an out-of-plane hyperfine coupling component $A_\perp = 130$ MHz [28], resulting in a total of 16 energy levels (Fig. 1a). The hyperfine coupling causes a shift in the ESR transition frequency that depends on the nuclear spin state, which is absent for Ti isotopes with $I = 0$ (Figs. 1b and c). The observation of multiple ESR peaks in a single sweep indicates that the nuclear spin state changes much faster than the averaging time of 3 seconds for each datapoint under conventional ESR settings. The peak heights are a measure of the population distribution of the various nuclear spin states: the nuclear spin resides longer in the $m_I = -7/2$ state as a result of nuclear spin pumping by inelastic electron scattering[27,28], where $m_I$ refers to the magnetic quantum number of the nuclear spin along the external magnetic field $B_z$. Being a time-averaged measurement, however, the ESR frequency sweep does not provide information on nuclear spin transition timescales.

In order to resolve changes between the nuclear spin states, we measured the ESR signal in the time domain instead of in the frequency domain. We switched off the STM feedback loop and sent a continuous RF signal to the tunnelling junction with a fixed frequency $f_{probe}$, resonant with a transition of the electron spin magnetic quantum number $m_S$ between states $|m_S, m_I\rangle = |\downarrow, -7/2\rangle$ and $|\uparrow, -7/2\rangle$, resulting in an additional ESR current $I_{ESR}$ only when $m_I = -7/2$, as illustrated in Fig. 1a. Figure 1d shows the current measured for several seconds, revealing stochastic switching between two discrete levels. The two levels become apparent in a histogram of the time trace as two Gaussian distributions separated by a current offset. In contrast, when we use an off-resonance frequency, the current is distributed as a single Gaussian (Supplementary Fig. 1a). We also performed reference measurements on a Ti isotope with $I = 0$ (Supplementary Figs. 1b and c), each revealing a single Gaussian distribution as well. We thus attribute the observed switching to quantum jumps between the probed state ($m_I = -7/2$) and any of the other seven nuclear spin states ($m_I \neq -7/2$).

Since the observed switching time is considerably longer than the averaging time (20 ms), the majority of datapoints constitute a single-shot readout of the nuclear spin. Note that the nuclear spin state is projected at a singular event sometime during the averaging time by an electron passing the adatom. The state is then read out repeatedly during the remaining time to integrate the current to reach a measurable signal strength. This is possible because, in the product state limit, reading out the electron spin state mostly leaves the nuclear spin unchanged after projection. We achieve readout fidelities of up to 98% for both the probed state being occupied and it being unoccupied (see Supplementary Information).

We can probe different nuclear spin states without changing $f_{probe}$ by adjusting the height of the magnetized probe tip, which exerts an additional magnetic field on the atom[29]. As we adjust the tip height (i.e., setpoint current), different values of $m_I$ become resonant with $f_{probe}$ (Fig. 2a). We measured current time traces at different tip heights (i.e., different total fields), resulting in a set of current histograms plotted together in Figs. 2b-f. Within the sweep range of the setpoint current, the histograms are found to feature a bimodal distribution in three windows around 3.2, 2.8, and 2.4 pA, matching the resonances of the nuclear spin states $m_I$ = −7/2, −5/2, and −3/2, respectively. While for $m_I$ = −7/2 the distribution is approximately 50-50 (Figs. 2e,f), the other states are found to be occupied much less than half of the time (Fig. 2c).

**Lifetime of the nuclear spin**

Analysis of the time traces enables us to extract the dwell times of the nuclear spin states. We label each point in the time trace as either in the probed state or not, using a threshold determined by the intersection point of the two Gaussian distributions (Fig. 1d). We measure individual dwell times $t_{dwell}$ for each occurrence of consecutive time spent in the probed state (Fig. 3a). Using an exponential fit, we find a lifetime $T_1^{CW}$ of about 100 ms for $m_I$ = −7/2 (Fig. 3a). We note that this is the nuclear spin lifetime under continuous-wave (CW) ESR driving and readout by the tunnelling current.

To compare $T_1^{CW}$ between different nuclear spin states, we repeated the experiment for different $f_{probe}$ values while leaving the tip height unchanged to avoid variations in current-induced nuclear spin pumping (Figs. 3b and c, see also Supplementary Fig. 2). We observe that for $f_{probe}$ being resonant with any of the $m_I \neq -7/2$ transitions, $T_1^{CW}$ is significantly shorter than for $m_I$ = −7/2. This may be partly attributed to the fact that selection rules allow these states to transition in both directions ($\Delta m_I = \pm 1$) whereas the $m_I$ = −7/2 state can only switch to $m_I$ = −5/2 ($\Delta m_I$ = +1). In addition, these experiments were performed in the presence of a spin-polarized current which constantly excites the system towards $m_I$ = −7/2.

However, neither of the above arguments can explain the fact that $T_1^{CW}$ keeps decreasing beyond $m_I$ = −5/2, suggesting that the spin pumping efficiency depends on the nuclear spin state. We believe this may be because the hybridization between neighbouring spin states depends on $m_I$ directly, which derives from the Clebsch-Gordan coefficients (see Supplementary Table 1). In addition, Fig. 3d shows how $T_1^{CW}$ changes as we vary the tip height around the resonance point for $m_I$ = −7/2. We find that $T_1^{CW}$ is increasing further up to ~300 ms as the system is detuned from resonance. This implies that the continuous ESR driving of the electron spin may affect $T_1^{CW}$ as well.

In order to find the intrinsic lifetime $T_1$ of the undisturbed nuclear spin and to identify its limiting relaxation mechanisms, we applied a pulsed readout scheme illustrated in Fig. 4a (see Methods). Here, we set the voltage (both DC and RF) to zero in between pulses. The probe pulses consisted of a similar DC + RF conditions as in the continuous-wave experiments mentioned above. We set the tip height such that $f_{probe}$ matches the resonance frequency for $m_I = -7/2$. As shown in Fig. 4b, the voltage pulses lead to sharp peaks in the current, each of which we can assign either the value 1 (for $m_I = -7/2$) or 0 ($m_I \neq -7/2$), using a threshold as introduced in Fig. 1d. For each pair of probe pulses A and B, there are four possible event scenarios, resulting in conditional probabilities $P(B|A)$ (Figs. 4c and d).

By changing the waiting time $\tau$ between the pulses, we then find the conditional probability as a function of $\tau$ (Fig. 4e). We take the unintentional pumping of the nuclear spin by the second probe pulse into account, correcting $P(1|1)$ to $P^*(1|1) = [P(1|1) - P(1|0)]/[1 - P(1|0)]$ (see Supplementary Information). $P^*(1|1)(\tau)$ is expected to decrease exponentially as a function of $\tau$ due to relaxation processes at a timescale of $T_1$. By fitting with an exponential function, we obtain $T_1 = 5.3 \pm 0.5$ seconds for $m_I = -7/2$. Similar measurements on an identical $^{49}$Ti atom yielded a value of $T_1 = 4.3 \pm 0.8$ seconds (Supplementary Fig. 3). This $T_1$ is seven orders of magnitude larger than the lifetime of the electron spin in the same atom (~100 ns)[16]. In addition, the extracted lifetime is an order of magnitude larger than $T_1^{CW}$ measured in the continuous-wave experiment (Figs. 1 and 2), confirming the hypothesis that the nuclear spin state is affected by continuous ESR driving and DC readout.

**Relaxation mechanism**

To develop an understanding on how DC spin pumping and ESR driving affect the nuclear spin, we repeat the pulsed experiment introduced in Fig. 4, but now with additional voltages in the waiting time to investigate perturbation of the nuclear spin (Figs. 5a-c) (see Methods). In a first variant we add a DC bias voltage during the waiting phase and fix the waiting time to 2 sec. The figure of merit is $P(1|1)$, now left uncorrected by $P(1|0)$ as we intentionally disturb nuclear spins during the waiting time (see Supplementary Information). In Fig. 5a we plot $P(1|1)$ against the bias voltage in the waiting phase. $P(1|1)$ is found to increase with positive voltage whereas it decreases with negative voltage.

We attribute this to dynamic nuclear spin polarization resulting from electron spin pumping, which is a combination of two processes depicted in Fig. 5d: flip-flop quantum jumps between a nuclear and electron spin due to hybridization as a result of the hyperfine coupling, and inelastic electron scattering. The former can happen both ways, while the latter is favoured in one direction due to the spin polarization of the tip. As a result, the two processes together pump the system towards $m_I = -7/2$ or $m_I = +7/2$, depending on the voltage polarity[27,28].

We note that, since Ti on the oxygen binding site displays a continuous-wave ESR signal only for positive voltages[30], it was previously not possible to observe nuclear spin pumping at negative bias voltage. With our pulsed readout scheme, however, we can now also confirm pumping in this bias regime. This points to an interesting aspect of our experiment compared to previous studies: we isolate the effect of the DC current from the ESR driving, the effect of which will be studied separately below.

In a second and third variant of the experiment, we send an RF signal during the waiting phase and fix the waiting time to 600 ms. In Figs. 5c and d, we change the frequency and RF amplitude, respectively, to see the effect of detuning and ESR driving strength on the conditional probability. During the probe pulses, $f_{probe}$ is kept at 12.75 GHz, corresponding to the $m_I = -7/2$ resonance. As we sweep the interim frequency, $P(1|1)$ dips down on resonance at around 12.75 GHz (Fig. 5b). This is an indication that the nuclear spin relaxes significantly faster when the electron spin is driven most efficiently. As shown in Fig. 5c, this effect increases with ESR driving amplitude but becomes weaker when the frequency is detuned.

This ESR-induced relaxation can be understood in terms of the availability of the relaxation pathway towards $m_I = -5/2$ via nuclear-electron spin flip-flops. Due to conservation laws, this process can only happen when the electron spin is in the $m_S = \uparrow$ state, whereas $m_S = \downarrow$ is the ground state. The nuclear spin at $m_I = -7/2$ can relax to $m_I = -5/2$ by a hyperfine flip-flop interaction, transferring $-1$ angular momentum to the electron spin (see Fig. 5d for schematic). Thus, the relaxation rate is amplified when $m_S = \uparrow$ state becomes more populated through ESR-driving.

We model the pulsed experiments using rate equation simulations. The model takes into account all interactions visualized in Fig. 5d, including the hybridization due to the hyperfine coupling[27] and an electron spin Rabi drive term. Interestingly, as discussed in the Supplementary Information, the simulation cannot quantitatively reproduce the experimental nuclear lifetime using realistic parameter values. Consequently, we choose to apply the model for qualitative comparison only. The model reproduces the DC pumping behaviour (Fig. 5e), and captures the observed dip in the frequency sweep (Fig. 5f) as well as the dependence on driving amplitude (Fig. 5g).

The pulsed experiments together with the simulations provide insight into the relaxation mechanisms limiting the intrinsic lifetime of the nuclear spin. We emphasize that we observed the relaxation by ESR driving to be strong in relation to any other relaxation source present. This indicates that the flip-flop interaction, which facilitates the relaxation by ESR, happens fast enough to be a dominating relaxation process whenever ESR continuously excites the electron spin. We note that the same flip-flop rate is present also when the electron spin is excited in ways other than ESR driving. In thermal equilibrium at 400 mK, electron scattering from the substrate significantly populates the $m_S = \uparrow$ state[31]. We therefore conclude that the nuclear-electron spin flip-flop relaxation channel is likely the main mechanism limiting the intrinsic lifetime.

As the flip-flop interaction results from a small in-plane component of the hyperfine coupling ($A_\parallel = 10$ MHz [25]), which in turn depends on the binding site[24], the above implies that the atomic-scale position is a crucial parameter to extend the lifetime of on-surface nuclear spins. Furthermore, the flip-flop relaxation channel is expected to be reduced at higher magnetic fields where the hybridization is minimized[3]. At that point, other relaxation sources may become relevant, such as spin-lattice coupling mediated by the nuclear quadrupole moment or magnetic Johnson noise from the bulk silver. Dipole-dipole flip-flops with the Mg and Ag nuclear spins can be considered negligible (see Supplementary Note 4).

In conclusion, we performed single-shot readout of an individual nuclear spin using ESR-STM, which was found to have a lifetime in the order of seconds. Furthermore, we shed light on the pumping channel by a local DC bias and relaxation channel by ESR driving. As the single-shot readout ESR-STM method presented here should work for any long-lived nuclear spin, the methodology may be transferable to different atomic or molecular spins in various platforms, such as semiconducting or insulating substrates[32,33]. Crucially, the nuclear spin lifetime we observe is longer than the rise time of current amplifiers conventionally used for STM, which enables direct readout and high-fidelity state initialization by projective measurement[34]. Potentially, this could even combine spin operations with STM tip movements between different atoms and open up possibilities to perform simultaneous coherent operations on extended atomic structures comprising multiple spins.

**Methods**

The experiments were performed in a commercial Unisoku USM-1300 STM. MgO/Ag sample and surface atoms were prepared with the same method described in Refs. [21] and [28]. The tungsten STM tip was prepared for ESR-STM measurements by first indenting it into the Ag until we obtained scanning image quality, then picking up approximately 30 Fe atoms. All ESR measurement are taken at a sample temperature of 0.4 K. STM topography was obtained with a constant current mode and at 1.6 K. All DC bias values are reported with respect to the sample. The pre-amplifier for the tunnelling current used for all measurements is the NF corp. SA608F2, with its internal analog 300 Hz low-pass filter active and the NF corp. LP5393 as a low noise DC power supply for the pre-amplifier.

For CW-mode measurement (Figs. 1-3), we used a Rhode & Schwarz SMA100B signal generator to deliver the RF signal to the STM junction. The DC voltage is generated by a Nanonis V4 Digital-to-Analog converter. During the ESR frequency sweeps, the RF signal was chopped at the frequency of 271 Hz. The ESR signal (difference between driven and non-driven) was detected at this frequency by a Stanford Instruments SR830m lock-in amplifier, resulting in an increased signal-to-noise ratio (SNR) compared to direct averaging of the current. For the time-trace measurement that resolve nuclear spin state switching, the driving RF voltage is not chopped and the current is instead recorded directly at a sampling time of 20 ms by a Nanonis V4 Analog-to-Digital converter (ADC). A wiring schematic of the components used for CW-mode measurement is shown in Supplementary Fig. 4a. When recording a CW-mode time trace, we have the RF signal generator running before turning off the tip-height feedback. This provides time for the thermal drift in tip height due to heating to reach an equilibrium to prevent a time-varying drift during the recording.

For pulse-mode measurement (Figs. 4 and 5), the voltage pulses were generated by a Keysight M8195a arbitrary waveform generator (AWG). A single channel was used to generate both the DC and RF voltages directly as one signal. A wiring schematic of the components used for pulse-mode measurement is shown in Supplementary Fig. 4b. In the pulse-mode the feedback is turned off first before running the AWG, but we start recording after tens of seconds to wait out the drift from heating. The drift settles faster compared to CW-mode due to the short duty cycle of RF pulses. Each section of a signal (probe A, waiting period, probe B, and separation period) is generated by repeating a shorter waveform with a length of 10 μs. It is justified to not have a phase-coherent RF pulse during the probe pulses, because the coherence time of the ESR-driven electron spin is only 300 ns at best[22]. For all pulsed experiments presented with τ < 2 seconds, the separation period has a length of 2 seconds. The separation period is equal to the waiting period for longer τ.

Pulsed experiments are recorded at the Nanonis ADC as full time traces with a sampling rate of 20 ms. A marker signal is sent by the AWG simultaneously with each probe pulse and is recorded at each ADC sample alongside the tunnelling current, in order to select the probe pulse samples from all other samples in the time trace for the data analysis. At least 40 ms probe pulse width was chosen to always leave a full 20 ms sample in the centre to avoid jitter artifacts at the edges of ADC samples. For details on probe sample selection and subsequent analysis, see Supplementary Information.

**Data availability**

The raw data generated in this study as well as the analysis and simulation code have been deposited in a Zenodo database under identifier https://doi.org/10.5281/zenodo.15518772.

**Acknowledgements**

This work was supported by the Dutch Research Council (NWO Vici Grant VI.C.182.016) and by the European Research Council (ERC Advanced Grant No. 101095574 "HYPSTER"). P.W. acknowledges funding from the Emmy Noether Programme of the DFG (WI5486/1-1) and the Daimler and Benz Foundation. L.M.V. acknowledges funding from the Alexander von Humboldt foundation.

**Author contributions**

E.W.S., J.L., P.W., and S.O. conceived the experiment. E.W.S and J.L. performed the experiment. E.W.S., J.L., and E.T. analysed the experimental data. H.G.V. and R.B. performed the calculations. E.W.S., J.L., H.G.V., R.B., A.J.K., L.M.V., P.W. and S.O. discussed the results. E.W.S., J.L., and S.O. wrote the manuscript, with input from all authors. S.O. supervised the project.

**Competing interests**

The authors declare no competing interests.

**Materials & Correspondence**

Correspondence and requests for materials should be addressed to Sander Otte or Jinwon Lee.

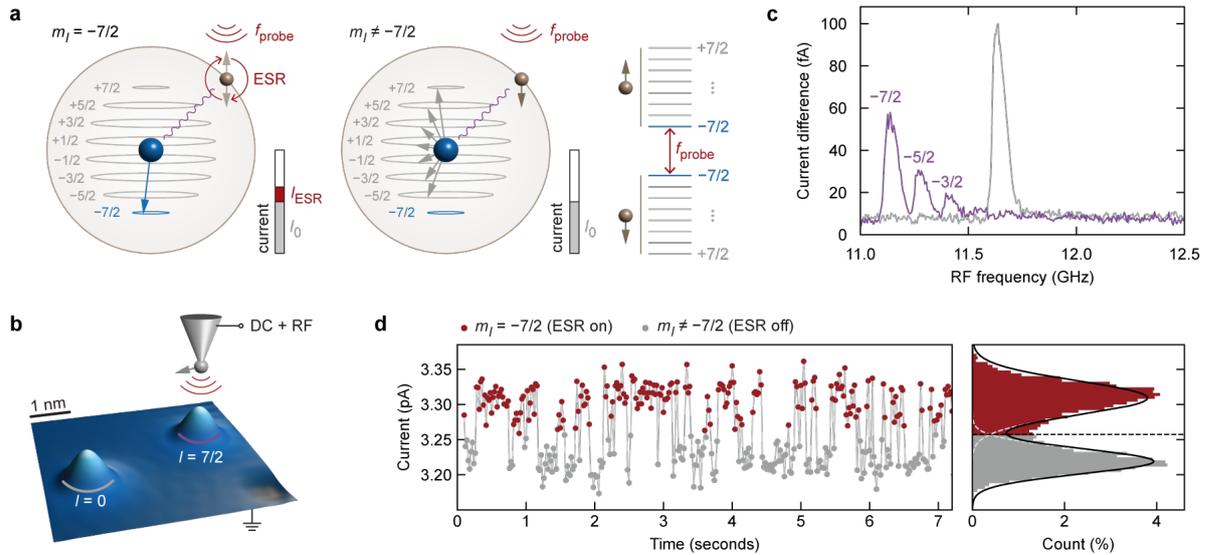

**Fig. 1 | Time-resolved switching of the nuclear spin state. a**, Diagram of the measurement scheme: the $I = 7/2$ nuclear spin (blue) of a $^{49}$Ti atom coupled to the $S = 1/2$ electron spin (brown) via the hyperfine interaction. Only when the nuclear spin is in the $m_I = -7/2$ state, does the applied RF signal at frequency $f_{probe}$ lead to a tunnelling current increase $I_{ESR}$. Right: spin energy diagram in the limit of a strong magnetic field, where the Zeeman splitting dominates and the eigenstates can be approximated by product states. **b**, STM topography (10 pA, 60 mV) of two Ti isotopes on MgO/Ag(100). **c**, ESR frequency sweeps (3.0 pA, 60 mV, $B_z = 1.35$ T, $V_{RF} = 17$ mV, averaging time 3 seconds, lock-in frequency 270 Hz) measured on the two Ti atoms in (b). $V_{RF}$ refers to the zero-to-peak RF voltage amplitude at the tunnel junction. **d**, Section of a time trace of the tunnelling current at fixed tip height (60 mV, averaging time 20 ms) with $f_{probe}$ corresponding to $m_I = -7/2$. The current histogram on the right is fitted with a two-Gaussian distribution (black line).

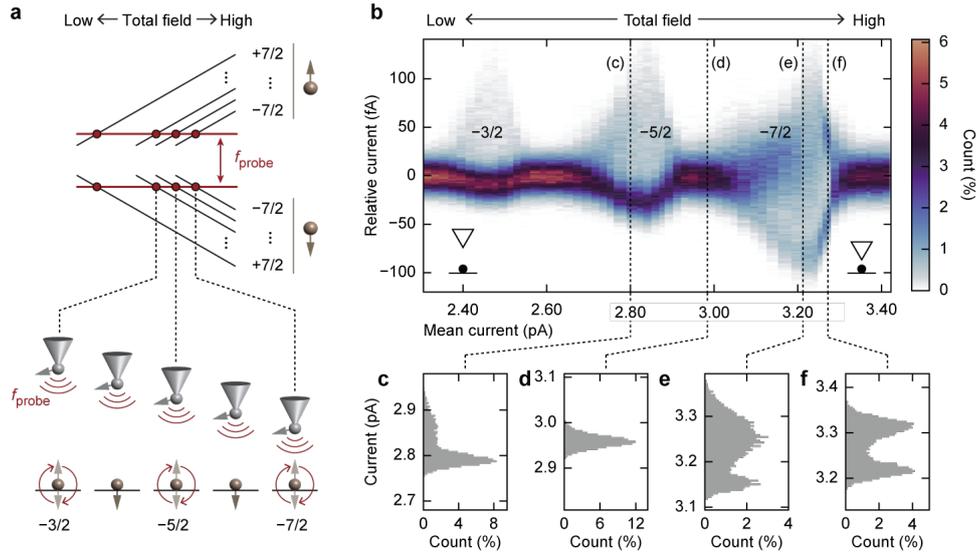

**Fig. 2 | Readout of different nuclear spin states. a**, Energy diagram of the spin states as a function of magnetic field (top) and schematic of the experiment shown in (b), indicating that at certain tip heights $f_{probe}$ matches the Zeeman splitting corresponding to one of $m_I$ states (bottom). **b**, Colour map of current histograms similar to Fig. 1d for different heights of the magnetic STM tip (bias voltage 60 mV). The horizontal axis represents the mean current for each time trace, which is a measure for the total magnetic field. **c-f**, Current histograms, corresponding to the labelled vertical dashed lines in (b).

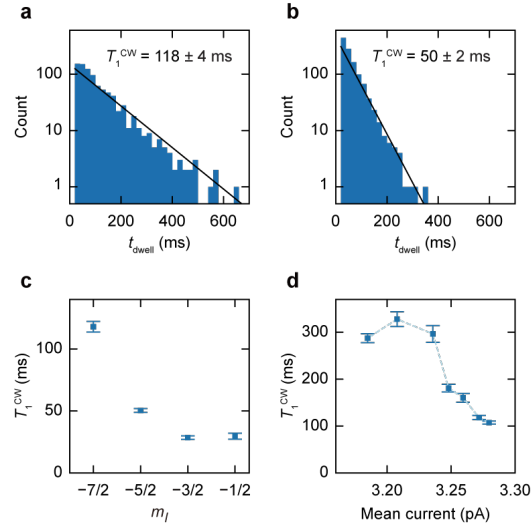

**Fig. 3 | Dwell time of different nuclear spin states. a** and **b**, Histograms of the distribution of the individual dwell times $t_{dwell}$ of $m_I = -7/2$ and $m_I = -5/2$, respectively. Fitting with an exponential function gives the characteristic dwell time $T_1^{CW}$. Error bars of $T_1^{CW}$ are the standard deviations in the exponential fit parameter. **c**, $T_1^{CW}$ for $m_I = -7/2, -5/2, -3/2, -1/2$, applying corresponding $f_{probe}$ at the same tip height. **d**, $T_1^{CW}$ of $m_I = -7/2$ at different tip heights around the resonance point.

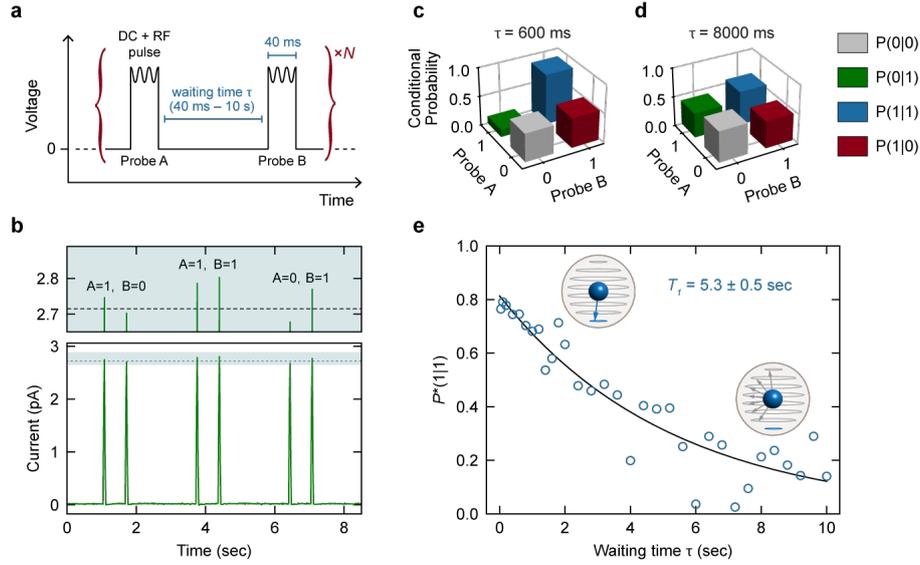

**Fig. 4 | Intrinsic nuclear spin lifetime with pulsed detection scheme. a**, Schematic of the pulse measurement scheme. One detection event consists of two probe pulses, with a variable waiting time τ in between with zero bias voltage. Each probe pulse consists of a DC voltage (70 mV) and an RF signal at $f_{probe}$ = 12.75 GHz, corresponding to $m_I$ = –7/2 ($B_z$ = 1.6 T, $V_{RF}$ = 22.6 mV). For each τ, we repeat between $N$ = 443 and $N$ = 704 detection events. **b**, Bottom: section of a measured current time trace (τ = 600 ms) showcasing three detection events. Top: zoom on the shaded area. For each event, the current of the probe pulses A and B indicates whether $m_I$ = –7/2 (1) or not (0), based on the current threshold (dashed line) determined from a histogram of all probe pulses. **c** and **d**, Conditional probabilities $P(B|A)$ for τ = 600 ms and τ = 8000 ms, respectively. **e**, Corrected conditional probability $P^*(1|1)$ as a function of τ. $P^*(1|1)$ represents the intrinsic decay of the nuclear spin. An exponential fit ($Ae^{-\tau/T_1}$, black curve) gives the intrinsic lifetime $T_1$ of the $m_I$ = –7/2 nuclear spin state ($T_1$ = 5.3 ± 0.5 sec, A = 0.81 ± 0.03).

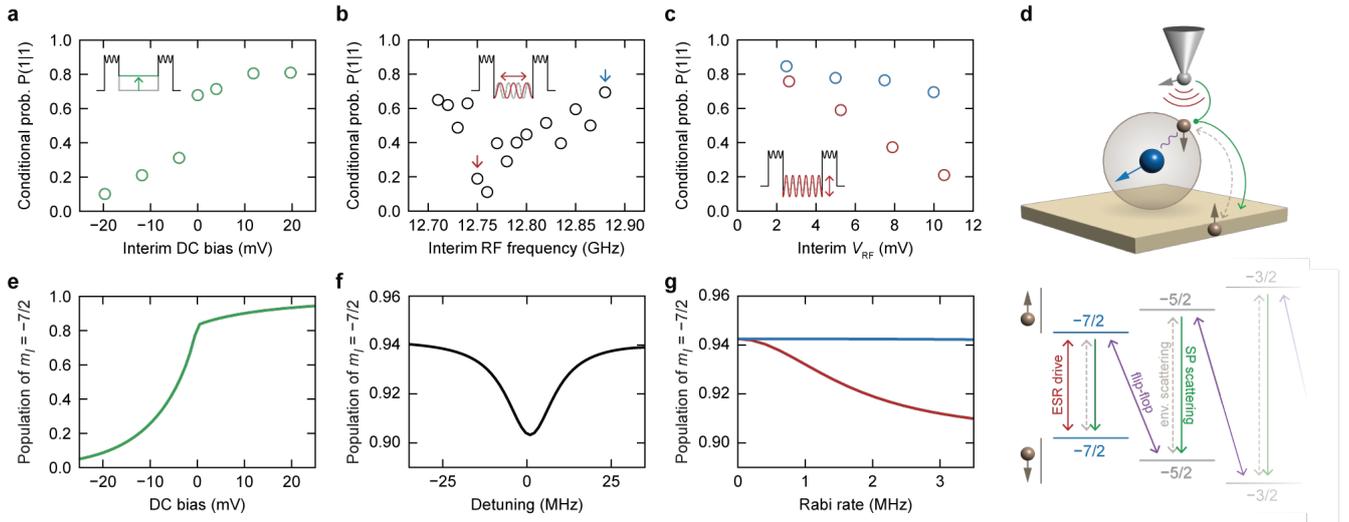

**Fig. 5 | DC or RF in waiting period. a**, Conditional probability $P(1|1)$ as a function of DC voltage during the waiting section between the two probe pulses (see inset schematic), with a fixed waiting time τ = 2 sec. **b**, $P(1|1)$ as a function of RF frequency in the waiting section (τ = 600 ms, $V_{RF}$ = 10.4 ± 0.3 mV). **c**, Conditional probability $P(1|1)$ as a function of RF voltage amplitude in the waiting section (interim $V_{RF}$), for two different frequencies highlighted in (b) (red: on-resonance frequency, blue: off-resonance frequency). **d**, Schematic of the possible transition dynamics between spin states: spin resonance drive of the electron spin (red), flip-flop quantum jump between nuclear and electronic spins mediated by hyperfine coupling (purple), environmental scattering of electronic spin (grey, dashed), electron spin scattering via spin-polarized STM tip (green). **e-g**, Rate equation simulations of the experiments performed in (a), (b), and (c) respectively. The model includes all processes indicated in (d) and implements them as transition rates between eigenstates. The populations are recorded after time evolution from the initial state $|\downarrow, -7/2\rangle$. Since the simulation does not quantitatively reproduce the experimental unperturbed nuclear lifetime, we chose simulated evolution times to be the same fraction of the simulated lifetime as the experimental ratio $\tau/T_1$ to capture a comparable moment in the decay process (see Supplementary Information).

# Supplementary Information for
# Single-shot readout of the nuclear spin of an on-surface atom

Evert W. Stolte[1,*], Jinwon Lee[1,*,†], Hester G. Vennema[1], Rik Broekhoven[1], Esther Teng[1], Allard J. Katan[1], Lukas M. Veldman[2], Philip Willke[3], Sander Otte[1,‡]

[1] Department of Quantum Nanoscience, Kavli Institute of Nanoscience, Delft University of Technology, Delft, The Netherlands
[2] Institute for Functional Matter and Quantum Technologies, University of Stuttgart, Stuttgart, Germany
[3] Physikalisches Institut, Karlsruhe Institute of Technology, Karlsruhe, Germany
* These authors contributed equally
† jinwon.lee@tudelft.nl
‡ a.f.otte@tudelft.nl

**Supplementary Note 1 | Analysis of pulsed experiment measurements**

For the probe pulse experiments, the current was measured as a continuous time trace, recording also the samples in between pulses as shown in Fig. 4b. Already from this time trace it can be observed that there are two current levels in the current, corresponding to the spin being in the $m_I = -7/2$ state (1) for the higher current or $m_I \neq -7/2$ (0) for the lower current. Conditional probabilities were extracted from this data set by systematically selecting the probe pulses from the raw time trace and then categorizing them by their nuclear spin state.

The sampling time of our recording Nanonis V4 analogue-to-digital converter (ADC) was set to 20 ms. To guarantee at least one full sample period per probe pulse, we sent probe pulses at least 40 ms wide. Since the probe pulses were not synchronised with the sampling, typically on either side of the pulse there was one sample containing a partial pulse (edge signals). Thus we obtained more than one data point per pulse. For example, with 40 ms pulses, we have three data points per pulse: one with the full sampling (green circles in Supplementary Fig. 5a) and two edge signals with a partial sampling (grey circles in Supplementary Fig. 5a). To select the full-sample signal, we sent a DC marker pulse along with our probe pulse and recorded the marker signal and tunnelling current simultaneously with our multi-channel ADC.

To guarantee a statistical validity for both the state labelling of individual pulses and the conditional probabilities calculated from these pulses, we need a large enough number of (i) events of probe-waiting-probe and (ii) current values that correspond to either nuclear spin states. To have sufficient events, we measured for at least 24 minutes for the shortest τ values, reaching up to 80 minutes for τ = 10 sec. In practice, time traces at a constant tip height can be recorded only for a limited amount of time before the drift in tip height changes the tunnelling current. Thus, we measured multiple time traces, restoring the tip height feedback to correct the drift before starting a new time trace, and combined them in the analysis. The drift in tip height during an individual time trace (8 minutes) is less than ~0.1 pm, equivalent to ~10 fA in current, which is smaller than $I_{ESR}$ that distinguishes the nuclear spin states in the probe pulses. For the measurements for τ ≥ 2 sec, we kept the separation time equal to the waiting time, where one probe pulse serves as the probe B for the former event and as the probe A for the latter event (Supplementary Fig. 5b). In this way, we could observe at least 443 events for each τ at a reasonable amount of measurement time. Note that for all experiments with a perturbing voltage in the waiting time (i.e. Fig. 5), the probes A and B were again analysed as distinguishable with the separation time fixed to 2 seconds at zero bias.

In order to have enough current values, we sent proportionally longer pulses – more samples per pulse – for longer τ. In this case, we used all current values with the full sample of the probe signal to construct a histogram and determine a current threshold, while we used only the first (probe B) and the last (probe A) value of currents of each pulse to calculate the conditional probability (Supplementary Fig. 5b). The current histogram of the full-sample signal (Supplementary Fig. 6a as an example, for τ = 8 sec) consists of the two Gaussian-distributed current levels corresponding to the nuclear spin states that we have seen in our CW measurements.

As we did in the CW measurement, we can define a threshold by fitting a current histogram with the sum of two Gaussians and taking their intersection point and label either (1) or (0) for each current. We construct a current histogram and obtain a threshold for each type of probe (probe A or B) individually unless they are indistinguishable in measurements with τ ≥ 2 sec. As shown in Supplementary Fig. 6b, almost all thresholds are within a narrow band ~20 fA wide, except the first three data points for the probe B. This is because of a certain settling time (Supplementary Fig. 6c) before the voltage in the ADC reaches a flat DC value, potentially due to the current pre-amplifier acting as a low pass filter. When the probe B follows shortly within this settling time, its signal adds the same signal amplitude on top of a different background, resulting in different current levels representing the same nuclear spin states.

After labelling the current for every probe, we obtained four conditional probabilities $P$(B|A) for each τ. As mentioned in the main text, we corrected for unintentional spin pumping in the probe B. Our figure of merit is the probability of the event that we observe the state (1) with both probe A and B. However, as illustrated in Supplementary Fig. 7, this inherently includes events where the state decays from (1) to (0) during the waiting time τ due to normal relaxation but is pumped back again to (1) with the probe B due to the DC voltage we need to readout the nuclear spin. To correct for this, we define $P_{10X}$ as the probability that (1) decays to (0) at least once during τ,

$$P_{10X} = P_{100} + P_{101}, \quad (1)$$

where $P_{100}$ and $P_{101}$ denote the probability of the event that we observe (1) with the probe A and it decays to (0) during τ, followed by the observation of (0) and (1) with the probe B, respectively. The latter case ($P_{101}$) corresponds to the nuclear spin pumping with the probe B. Only if we send zero voltage during τ, $P_{101}$ is expressed as,

$$P_{101} = P_{10X} \times P(1|0). \quad (2)$$

In addition, by the definition, $P(1|1) = P^*(1|1) + P_{101}$, and $P(0|1) = P_{100}$. Using Eqs. (1) and (2),

$$P(0|1) = P_{10X} - P_{101} = P_{10X} \times [1 - P(1|0)], \quad (3)$$

or

$$P_{10X} = \frac{P(0|1)}{1 - P(1|0)}. \quad (4)$$

By the definition,

$$P^*(1|1) = 1 - P_{10X} = \frac{P(1|1) - P(1|0)}{1 - P(1|0)}. \quad (5)$$

In a limit of τ → ∞, the event of observing (1) with the probe B is independent on the state measured with the probe A [$P(1|1) = P(1|0)$] (Supplementary Figs. 7c and d), yielding $P^*(1|1)$ converges to zero.

This correction is valid only if we do not send any voltage during τ. If we intentionally interrupt the nuclear spin during τ, Eq. (2) is not valid, which precludes the correction described above. This is why we use uncorrected $P(1|1)$ in Fig. 5 in the main text. Nevertheless, $P(1|1)$ still retains qualitative information about how likely the nuclear spin tends to remain in the (1) state. This allows for qualitative comparison between the experimental data and simulation results in Fig. 5.

There are also two assumptions made in the $P(1|1)$ correction described above, both of which we can justify experimentally. The first assumption is that (0) → (1) pumping events are mostly caused by the probe pulse and generally not by anything happening during the waiting time. If there would be a more 'intrinsic' mechanism of excitation to the $m_I = -7/2$ state during the waiting time, there must be a timescale associated with this process. This should then be reflected in our measured $P(1|0)$ as a function of waiting time τ. Instead, we find $P(1|0)(τ)$ to be constant (see Supplementary Fig. 7d). So, we conclude that pumping events are generally caused by a factor that does not change with the waiting time: the probe pulse.

The second assumption is that the nuclear spin always finds itself in a state from where it can be pumped up to $m_I = -7/2$. This might not be the case in general, since the $-7/2$ state can only be reached from the $-5/2$ state and there are 6 more nuclear spin projections. When measuring the probability $P(1|0)$, the initial state after the first probe pulse could be any of the states that fall under (0). When falling down from (1) during a $P(1|1)$ measurement, however, it is likely that the fallen state is $m_I = -5/2$ specifically and not any general state within (0). We want to correct $P(1|1)$ with $P(1|0)$ to account for pumping after falling, but $P(1|0)$ cannot be assumed to be representative of the probability to pump up soon after falling down from (1) due to this difference in initial state. So, $P(1|0)$ is a priori not the correct probability to use. However, we can take $P(1|0)$ as a good approximation if it can be shown that $m_I = -5/2$ is often occupied compared to other $m_I$ in the general (0) state. This is what we will demonstrate below.

Since we only need to investigate the $-5/2$ occupation during the probe pulse with continuous readout when the (0) → (1) pumping events happen, we can analyse the CW current time traces that probe individual $m_I$ states. From the magnitudes of the fitted $T_1^{CW}$ for the different states (Fig. 3c), we can see that the system is most likely to spend time in the $-5/2$ state if it is not in the $-7/2$ state. For $m_I = -5/2$ we find $T_1^{CW} = 50 ± 2$ ms, while we get $28 ± 1$ ms and $30 ± 2$ ms for $m_I = -3/2$ and $-1/2$, respectively.

Considering that our probe pulses are 40 ms wide, which is longer than the above characteristic $T_1^{CW}$ in both the $-3/2$ and $-1/2$ states, we expect that the nuclear spin is often pumped up to at least the $-5/2$ state during a probe pulse. In other words, if the nuclear spin is neither occupying the $-7/2$ nor the $-5/2$ state during the second probe pulse, it will soon be occupying either of them within the 40 ms probe pulse width. In fact, the time needed to pump to $-5/2$ might be even shorter while probing $-7/2$ compared to time traces probing the $-3/2$ and $-1/2$ states directly, which we did to get the above $T_1^{CW}$, because ESR driving inhibits the spin pumping of the electron spin at the probed state.

Overall, this means that we can justify both our assumptions and our correction of $P(1|1)$ to $P^*(1|1)$ can be applied. This correction is important for a quantitative analysis, in particular to apply proper fitting to get the correct $T_1$. Using $P^*(1|1)$, we get a result for $T_1$ that is, in principle, independent of probe pulses properties like the DC voltage pulse height.

**Supplementary Note 2 | Readout fidelity and modelling of the distribution of sample currents**

Following Ref. 1, we calculate the readout fidelity (also referred to as measurement fidelity) using a model of the separate sample current distributions $N_1(I)$ and $N_0(I)$ for when the nuclear spin occupies the probed spin state (1: $m_I = -7/2$) and the other spin states (0: $m_I \neq -7/2$), respectively. Knowing these separate distributions, the readout fidelities $F_0$ and $F_1$ are then defined as the fractions of the distributions on the correct side of the chosen current threshold $I_{\text{thr}}$:

$$F_0 = \int_{-\infty}^{I_{\text{thr}}} N_0(I) dI, \qquad (6)$$

$$F_1 = \int_{I_{\text{thr}}}^{\infty} N_1(I) dI. \qquad (7)$$

Experimentally, we have access only to the total sample current distribution from both distributions combined. As such, before calculating the fidelity, we need to model the separate sample current distributions and fit this model to the experiment. For all data analysis presented in the main paper, we opted to model our measurements with an analytical model where $N_0(I)$ and $N_1(I)$ are both Gaussian distributions. The resulting fit with the experimental current distribution of one particular time trace is shown in Fig. 1d. This model is justified because STM current noise is generally Gaussian-distributed.

We can find an optimal value for the threshold current where the maximum number of samples will be labelled correctly. Naturally, this value is the intersection point between the two Gaussians, resulting in readout fidelities $F_0$ = 98% and $F_1$ = 98% for the time trace showcased in Fig. 1d, which was chosen with an optimal ESR drive. Other time traces generally show lower fidelities, in particular due to an increased current noise for the probed nuclear spin state when driven detuned (see, for example, Fig. 2e). Across all analyzed CW and lifetime pulse-mode time traces reading out the $m_I = -7/2$ state, we find a mean fidelity of $F_0$ = 91 ± 5 % and $F_1$ = 90 ± 4 %.

**Supplementary Note 3 | Rate equation model for nuclear spins**

The model Hamiltonian for our system consists of Zeeman interactions for both the electron spin and the nuclear spin, the hyperfine interaction between the nuclear and electron spin, and the quadrupole interaction of the nuclear spin. For this system we have a total electronic spin $S = 1/2$ and a total nuclear spin for the $^{49}$Ti isotope $I = 7/2$, with spin projection operators $\hat{S}_i$ and $\hat{I}_i$, respectively, for the three axes $i = x, y, z$. The Hamiltonian reads as:

$$H = \sum_{i=x,y,z} \left( g_{e,i} \mu_B \hat{S}_i (B_{\text{ext},i} + B_{\text{tip},i}) + g_{N,i} \mu_N \hat{I}_i (B_{\text{ext},i} + B_{\text{tip},i}) + A_i \hat{I}_i \hat{S}_i + Q_i \hat{I}_i^2 \right). \qquad (8)$$

Here, **g**$_e$ is the anisotropic g-factor of the electron spin, [1.61, 1.61, 0.67] [2], $\mu_B$ is the Bohr magneton, $g_N$ is the nuclear g-factor of -0.315 [3], $\mu_N$ is the nuclear magneton, **A** is the anisotropic hyperfine constant [10, 10, 130] MHz and **Q** the quadrupole moment of [1.5, 1.5, -3] MHz [4]. The external magnetic field is along the out-of-plane direction $z$, which defines the quantization axis of the system. We assume the magnetic field of the tip **B**$_{\text{tip}}$ to be mostly aligned with the out-of-plane external magnetic field **B**$_{\text{ext}}$, to which we add a small in-plane component to **B**$_{\text{tip}}$, necessary for ESR driving. To match the observed resonance frequencies, we use $B_{ext}$ = 1.35 T along $z$ and a $B_{tip}$ = 0.197 T with a 2° rotation angle with respect to the out-of-plane direction.

Diagonalization of the Hamiltonian yields the $(2S+1)(2I+1)$ = 16 eigenstates and corresponding energies. As the applied field is large compared to the hyperfine interaction, the eigenstates $|n\rangle$ are well described by Zeeman product states (see Supplementary Table 1). Nonetheless, some hybridization remains present allowing for flip-flop interactions between those states[5].

The population $P_n$ of each eigenstate can be determined by calculating the transition rates $r_{nm}$ in and out of the eigenstate. The change in population of each eigenstate is then determined by the following first order differential equation[6]:

$$\frac{dP_n(t)}{dt} = \sum_m r_{nm} P_m(t) - r_{mn} P_n(t), \tag{9}$$

where

$$r_{nm} = r_{nm}^{tt} + r_{nm}^{ss} + r_{nm}^{ts} + r_{nm}^{st} + r_{nm}^{ESR}. \tag{10}$$

The first four of these are based on Kondo scattering (electron scattering). We allow interaction for the electron spin with the sample electron bath ($r_{nm}^{ss}$), with electrons scattering from the tip to the sample and vice versa ($r_{nm}^{ts}$, $r_{nm}^{st}$), and we also take into account tip electron bath scattering ($r_{nm}^{tt}$). These rates are described as,

$$r_{nm}^{tt} = \int f(\varepsilon)(1-f(\varepsilon))\varrho_t^2 \sum_{\sigma,\sigma'} J_t^2 |\langle n,\sigma|\widehat{\boldsymbol{S}}\cdot\widehat{\boldsymbol{\sigma}}|m,\sigma'\rangle|^2 \, d\varepsilon, \tag{11}$$

$$r_{nm}^{ss} = \int f(\varepsilon)(1-f(\varepsilon))\varrho_s^2 \sum_{\sigma,\sigma'} J_s^2 |\langle n,\sigma|\widehat{\boldsymbol{S}}\cdot\widehat{\boldsymbol{\sigma}}|m,\sigma'\rangle|^2 \, d\varepsilon, \tag{12}$$

$$r_{nm}^{ts} = \int f(\varepsilon)(1-f(\varepsilon+eV))\varrho_t\varrho_s \sum_{\sigma,\sigma'} \rho_\sigma^t J_t J_s |\langle n,\sigma|\widehat{\boldsymbol{S}}\cdot\widehat{\boldsymbol{\sigma}}|m,\sigma'\rangle|^2 \, d\varepsilon, \tag{13}$$

$$r_{nm}^{st} = \int f(\varepsilon+eV)(1-f(\varepsilon))\varrho_s\varrho_t \sum_{\sigma,\sigma'} \rho_{\sigma'}^t J_s J_t |\langle n,\sigma|\widehat{\boldsymbol{S}}\cdot\widehat{\boldsymbol{\sigma}}|m,\sigma'\rangle|^2 \, d\varepsilon, \tag{14}$$

where $f(\varepsilon)$ is the Fermi Dirac function with $V$ the bias voltage applied to the tip with respect to the sample; $\varrho_t$ and $\varrho_s$ are the bath electron densities for the tip and sample, respectively. The summation inside each integrand describes all possible scattering processes by taking the transition element between the initial ($n$, $\sigma$) and final ($m$, $\sigma'$) states (each product states of an eigenstate of the Hamiltonian and a spin state of the scattering electron spin $\sigma$) due to flip-flop interactions between $\widehat{\boldsymbol{S}}$ and $\widehat{\boldsymbol{\sigma}}$. The different scattering processes are distinguished by the polarization direction of the tip, and the various corresponding effective (spin-spin) couplings $J$ between the baths (s, t) and the surface (electron) spin. The magnetic polarization of scattering electrons is described by $\rho_\sigma^t = 1/2 + \eta\sigma$. Here, we have taken the tip polarization $\eta$ to be completely polarized (i.e. equal to 1).

Additionally, we add the transition rate for ESR interaction as an averaged transition rate[5]:

$$r_{nm}^{ESR} = \frac{1}{2} 4\pi^2 \Omega^2 |\langle n|\hat{S}_x|m\rangle|^2 \frac{T_2}{4\pi^2(\delta f_{RF} - |E_n - E_m|)^2 T_2^2 + 1} \tag{15}$$

with $\Omega$ as Rabi rate or driving power and $\delta$ as detuning factor of the driving frequency $f_{RF}$. The transition matrix element between initial and final state with operator $\hat{S}_x$ gives the electron spin flips. Furthermore, $T_2$ is derived from the matrix of Kondo transition rates (i.e. Eq. (10) with $r_{nm}^{ESR} = 0$). Following Ref. 7, we sum for each transition the off-diagonal rates and take the inverse to arrive at a decoherence time.

$$\left.\frac{1}{T_2}\right|_{nm} = \frac{1}{2}\sum_{n'}(r_{nn'} + r_{mn'}). \qquad (16)$$

Here, for the driven ESR transition between eigenstates $|7\rangle$ and $|8\rangle$, approximately corresponding to $|-7/2,\downarrow\rangle$ and $|-7/2,\uparrow\rangle$ respectively (Supplementary Table 1), we find $T_2$ = 110 ns.

From the transition rate matrix, we can calculate an intrinsic lifetime for each eigenstate[6]. Then, looking at the full subspace for $m_I = -7/2$, consisting of the two eigenstates $|7\rangle$ and $|8\rangle$, we get to a value for the intrinsic lifetime of this nuclear spin state. Thus,

$$\tau_{7,8} = \left(\sum_{i\neq 7,8} r_{7,i} + r_{8,i}\right)^{-1}. \qquad (17)$$

To reproduce the intrinsic nuclear spin lifetime measured in our experiments, we investigated the lifetime as a function of the coupling strengths of the electron spin to the surroundings: $J_s$ and $J_t$, which are combined with $\varrho_s$ or $\varrho_t$ as unitless parameters. Since the interaction between the spins in the atom and electrons in the substrate is the dominant factor in limiting electronic spin lifetime, our starting point is to use $J_s\varrho_s = 10^{-2}$ to match the experimental value of this lifetime of 100 ns [8]. The coupling $J_t\varrho_t$ is experimentally variable and usually $J_s\varrho_s > J_t\varrho_t$, because the tip bath is further away from the atomic electron compared to the surface bath.

Interestingly, it seems impossible to achieve the nuclear spin lifetime in the order of seconds that was measured in our experiments by using $J_s\varrho_s = 10^{-2}$. Even completely decoupling the tip ($J_t\varrho_t = 0$) would result in a lifetime of tens of ms as shown in Supplementary Fig. 8. When taking $J_s\varrho_s = 10^{-3}$, it would be possible to get a lifetime in the order of seconds (Supplementary Fig. 8) while still obeying $J_s\varrho_s > J_t\varrho_t$. However, we must note that this low coupling of the electron spin would give rise to an unrealistic lifetime of the electron spin itself of 10 μs. Therefore, we use $J_s\varrho_s = 10^{-2}$ and limit ourselves to shorter intrinsic lifetimes of the nuclear spin state but correct lifetime of the electron spin. In addition, we chose $J_t\varrho_t = 10^{-4}$, which gave rise to appropriate spin pumping by DC voltage. In this model, we have deemed direct relaxation channels of the nuclear spin to be negligible, due to the readily high transition rate coming from electron spin coupling. This seems appropriate according to our theoretical investigation of the relevance of nuclear spin relaxation channels (Supplementary Note 4).

Summarising the above: we find a quantitative discrepancy between the calculated nuclear spin lifetime and the experimentally observed value. Therefore, we continue our theoretical investigation qualitatively and compensate for the discrepancy in our comparison with the experiment, which we will describe in detail below. A quantitative modelling of the lifetime is beyond the scope of this work.

The transition rate matrix is time-independent. Therefore, we can solve the time evolution by defining a linear set of equations. We determine the stationary eigenvectors $v_n$ and eigenvalues $\lambda_n$ of the transition rate matrix and calculate the populations $P(t)$ in time for each state:

$$P(t) = \sum_n \langle P(t=0)|v_n\rangle v_n e^{\lambda_n t}. \qquad (18)$$

We define a logarithmically spaced list of points in time to have enough sampling for the high frequency interactions and not oversample the low frequency interactions. The result is an evolution of the populations in time from an initial population distribution for one set of parameters that define the transition rates. Supplementary Fig. 9 shows an example of such a free evolution in time that represents the decay of the $m_I = -7/2$ state.

To simulate the influence on the relaxation processes of the DC and RF signals, we look at the decay of populations under varying conditions. For each simulation, we initialize the system to $P_{|\downarrow,-7/2\rangle}(t = 0) = 1$. Then we calculate the time evolution of the populations (from ns to sec). To compare with the experimental format, which has a fixed waiting time $\tau_{exp}$, we select one slice of the populations at a time value $\tau_{sim}$ from each time evolution. Since the simulation does not reproduce the experimental nuclear lifetime, but we want to capture a comparable moment in the decay process, we choose $\tau_{sim}$ such that $\tau_{sim}/T_{1,sim}$ is the same ratio as $\tau_{exp}/T_{1,exp}$ in the experiment (2/5 for DC and 0.6/5 for RF during the waiting time).

Comparing with the observable $P(1|1)$ in our experiment, the figure of merit is the total population in $m_I = -7/2$, which is the sum of $P_{|\downarrow,-7/2\rangle}$ and $P_{|\uparrow,-7/2\rangle}$. This is the result plotted in Fig. 5 against the DC voltage ($V$), Rabi rate ($\Omega$) or detuning factor ($\delta$). As demonstrated in these plots, we have been able to qualitatively reproduce the effects of the added signals in the waiting time on the population of the nuclear spin state.

**Supplementary Note 4 | Investigation of other nuclear spin relaxation channels.**

We argue in the main text, based on experimental results, that the nuclear spin lifetime of our $^{49}$Ti adatom is most likely limited by the degree of hybridization between the nuclear spin and the accompanying electron spin in the same atom. In this section, we will further support our reasoning with a theoretical investigation of other potential contributions to nuclear spin relaxation. We will discuss the following relaxation sources: (a) dipole-dipole coupling to other nuclear spins, (b) hyperfine coupling with bulk electron spins in silver, (c) charge fluctuations, (d) magnetic Johnson noise, and (e) phonon-mediated relaxation.

*(a) Dipole-dipole coupling to other nuclear spins*

One potential relaxation channel for the $^{49}$Ti nuclear spin might be via magnetic dipole-dipole coupling to nuclear spins in the MgO layer and the bulk silver underneath. Silver consists solely of isotopes with nuclear spin $I_{Ag} = 1/2$. The isotope $^{25}$Mg has nuclear spin $I_{Mg} = 5/2$ with a natural abundance of 10%, meaning there is a 34% chance of the $^{49}$Ti atom neighbouring at least one $^{25}$Mg. Oxygen also has a stable isotope with finite nuclear spin, but this isotope has a natural abundance of only 0.04%.

Relaxation via a direct flip-flop between two nuclear spins, a similar channel to the dominating hyperfine-mediated flip-flop with the electron spin, is highly suppressed because of the large energy detuning between the spins compared to the weak dipole coupling. This can be illustrated with a simplified toy-model Hamiltonian with two nuclear spins A and B, both $I^A$, $I^B = 1/2$, with an energy detuning $\Delta$ and an off-diagonal term $D$ from the dipole-dipole coupling:

$$\hat{H} = \Delta \hat{I}_z^A + D\left(\hat{I}_+^A \hat{I}_-^B + \hat{I}_-^A \hat{I}_+^B\right),$$

where $\hat{I}^A_z$ is a spin projection operator on the quantization axis for spin A, and $\hat{I}^A_+$, $\hat{I}^A_-$, $\hat{I}^B_+$, and $\hat{I}^B_-$ are the spin ladder operators for the two spins. Spin A is analogous to the $^{49}$Ti nuclear spin and spin B to a nuclear spin in the bulk. This Hamiltonian has four eigenstates that approach the spin product states in the regime of large detuning $\Delta$. In particular, one of the eigenstates has spin A in an excited state:

$$|E\rangle = \frac{1}{C}\left(\Delta + \sqrt{\Delta^2 + D^2}\right)|\uparrow,\downarrow\rangle + \frac{1}{C}|\downarrow,\uparrow\rangle \approx |\uparrow,\downarrow\rangle,$$

$$\text{with } C^2 = 2 + 2\frac{\Delta^2}{D^2} + 2\sqrt{\frac{\Delta^4}{D^4} + \frac{\Delta^2}{D^2}}.$$

Applying a Fermi golden rule argument, the rate of the flip-flop quantum jump to relax spin A – the inverse of the decay time $T_{flip}(\Delta)$ – scales with the probability $P_{flip}$ of finding $|E\rangle$ in the spin product state $|\downarrow,\uparrow\rangle$ upon projective measurement (by the environment). This probability is non-zero because of the hybridization and can be calculated from our eigenstate as

$$P_{flip} = |\langle\downarrow,\uparrow|E\rangle|^2 = \frac{1}{|C|^2} \sim \frac{1}{T_{flip}}.$$

This scaling can be combined with the decay time at zero detuning $T_{flip}(\Delta=0)$ to get $T_{flip}$ at any detuning. $T_{flip}(\Delta=0)$ is approximately the coherent flip flop time $1/D$, with $D$ in units Hz. Furthermore, in the regime of large detuning compared to the dipole coupling ($\Delta \gg D$), we can approximate $|C|^2$ by only keeping the terms quadratic in ($\Delta/D$), leading to the result

$$T_{flip} \approx T_{flip}(\Delta = 0) \cdot 4\left(\frac{\Delta}{D}\right)^2 = 4\frac{\Delta^2}{D^3}.$$

The $^{25}$Mg nuclear spin will often be closest to the $^{49}$Ti nuclear spin, resulting in the largest dipole-dipole coupling. The dipole-dipole flip-flop matrix element is given by

$$D = \frac{\mu_0 \gamma_{Ti} \gamma_{Mg} \hbar^2}{8\pi d^3} \cdot (3\cos^2\theta - 1),$$

with $\mu_0$ the vacuum magnetic permeability, $\gamma_{Ti}$ and $\gamma_{Mg}$ the gyromagnetic ratios for $^{49}$Ti and $^{25}$Mg nuclear spins respectively and Planck's constant $\hbar = h/2\pi$. We estimate the distance $d$ from the Ti nucleus to the closest Mg nucleus in the MgO layer to be 0.2 nm [9], with $\theta$ = 45° the angle with the quantization axis. Filling in these parameters, we find a dipole-dipole flip-flop coupling $D$ = 26 Hz. The detuning is mainly set by the hyperfine coupling, which is present only for $^{49}$Ti. For $m_I = -7/2$ we estimate a transition energy of ~49 MHz, which also includes nuclear Zeeman splitting at 1.6 T and the quadrupole moment. $^{25}$Mg has a Zeeman splitting of only ~4 MHz, thus $\Delta \approx 45$ MHz. The resulting estimated decay time from nuclear spin flip-flops $T_{flip}$ is then 4 x 10$^{11}$ seconds. Therefore, we deem this relaxation channel negligible.

We also considered flip-flops with other $^{49}$Ti on oxygen binding sites of MgO. Note that none were present on the same approx. 50 nm wide MgO island in our experiment, but a 20 nm separation distance would result in a reasonable $1/D$ = 12 hours. These nuclear spins of the same species are still detuned, however, because the magnetic field from the STM tip adds Zeeman splitting only for the spin under study[10]. A typical tip field strength of 20 mT corresponds to $\Delta$ = 302 kHz, which is enough to suppress this relaxation channel, too.

### (b) Hyperfine coupling with bulk electron spins in silver

Another less likely candidates for limiting the relaxation time is direct hyperfine coupling to electronic spins in the silver metallic bath[11] (Fermi-contact). There is too little wavefunction overlap between the nucleus and the silver electrons. With a mean free path of silver electrons larger than microns below 1 K [12], we estimate a Fermi-contact interaction that is at least 12 orders of magnitude smaller than the Ti electron hyperfine coupling (ratio of wavefunction volumes), resulting in a coupling of less than 0.1 mHz.

### (c) Charge distribution fluctuations through orbital excitations

Infrequent and short-lived fluctuations in the charge distribution of the Ti adatom could play a role in relaxation by modifying the hyperfine coupling and thus the hybridization between the $^{49}$Ti electron and nuclear spin. In particular, these fluctuations could be caused by thermal orbital excitations. Ti on the oxygen binding site does have an orbital excitation around 80 mV[13]. With our operating temperature of 0.4 K, however, thermal excitations into this orbital are expected to be negligible, making also this nuclear spin relaxation pathway unlikely.

### (d) Magnetic Johnson noise

A more significant contribution comes from magnetic Johnson noise originating from thermal ballistic electron transport in the silver bulk[7,14]. This magnetic noise can drive nuclear spin transitions, leading to spin relaxation. Extrapolating from experimental results on NV centers in close proximity to silver[14] — accounting for the different g-factor, temperature, and distance between the spin and the bulk — results in an estimated nuclear spin lifetime on the order of hours, if this relaxation channel would be limiting.

### (e) Phonon-mediated relaxation

Another potentially large contribution could be spin-lattice relaxation via inelastic Raman-like processes[15]. Phonons can couple to the nuclear spin in different ways. (i) Vibrations in the lattice move the nucleus through the gradient magnetic dipole field from Mg and Ag nuclear spins in the bulk or other nearby magnetic adatom on the surface, which can drive spin transitions. (ii) Moving point charges around the Ti nucleus modulate the electric field gradient at the position of the nucleus, which couples to its quadrupole moment. (iii) Moving point charges modulate the hyperfine coupling. This could also change the hyperfine-mediated flip flop rate.

In these spin-lattice relaxation mechanisms, one phonon from a THz band scatters on the nuclear spin and returns to the same band with a slightly different energy. This is limited, however, by the thermal occupation of the phonon modes. Spin-phonon scattering rates fall off quickly with temperature far below the Debye temperature[15], which we estimate to be 1 x10$^3$ K in our MgO thin film[16]. Therefore, spin-lattice relaxation is likely not a factor relative to the hyperfine-mediated flip-flop with the Ti electron spin at our operating the magnetic field. For reference, the nuclear spin of $^{25}$Mg in MgO, which has a similar quadrupole moment as $^{49}$Ti and is also positioned next to oxygen in the lattice, has a $T_1$ of 49 seconds at room temperature[17].

Since the spin-lattice relaxation does not depend on magnetic field strength, we hypothesize that spin-lattice relaxation could be more relevant than the hyperfine-mediated flip flop relaxation channel at high magnetic fields. Of the three coupling mechanisms, quadrupole-mediated coupling is often the strongest[18] and can thus be expected to be the largest contribution in this regime.

We find that all sources of nuclear spin relaxation discussed above have a characteristic lifetime that is orders of magnitude longer than lifetime of 5.3 seconds we found experimentally. Supported further by the observed enhancement of relaxation due to ESR driving, which is discussed in the main text, we conclude that the nuclear spin lifetime of $^{49}$Ti in our experiment is limited by the hyperfine-mediated flip-flop between the nuclear spin and the Ti electron spin. In the absence of this relaxation channel, the next-biggest contributions to relaxation might be magnetic Johnson noise or spin-lattice coupling.

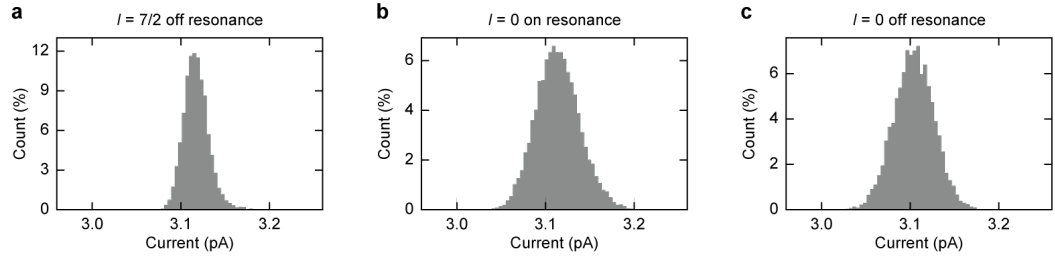

**Supplementary Fig. 1 | Current histograms of time traces. a**, Current histogram with off-resonance frequency on $^{49}$Ti. **b** and **c**, Current histogram with on-resonance and off-resonance frequency on $^{48}$Ti ($I$ = 0), respectively. All histograms are unimodal, which indicates stochastic switching in current is absent.

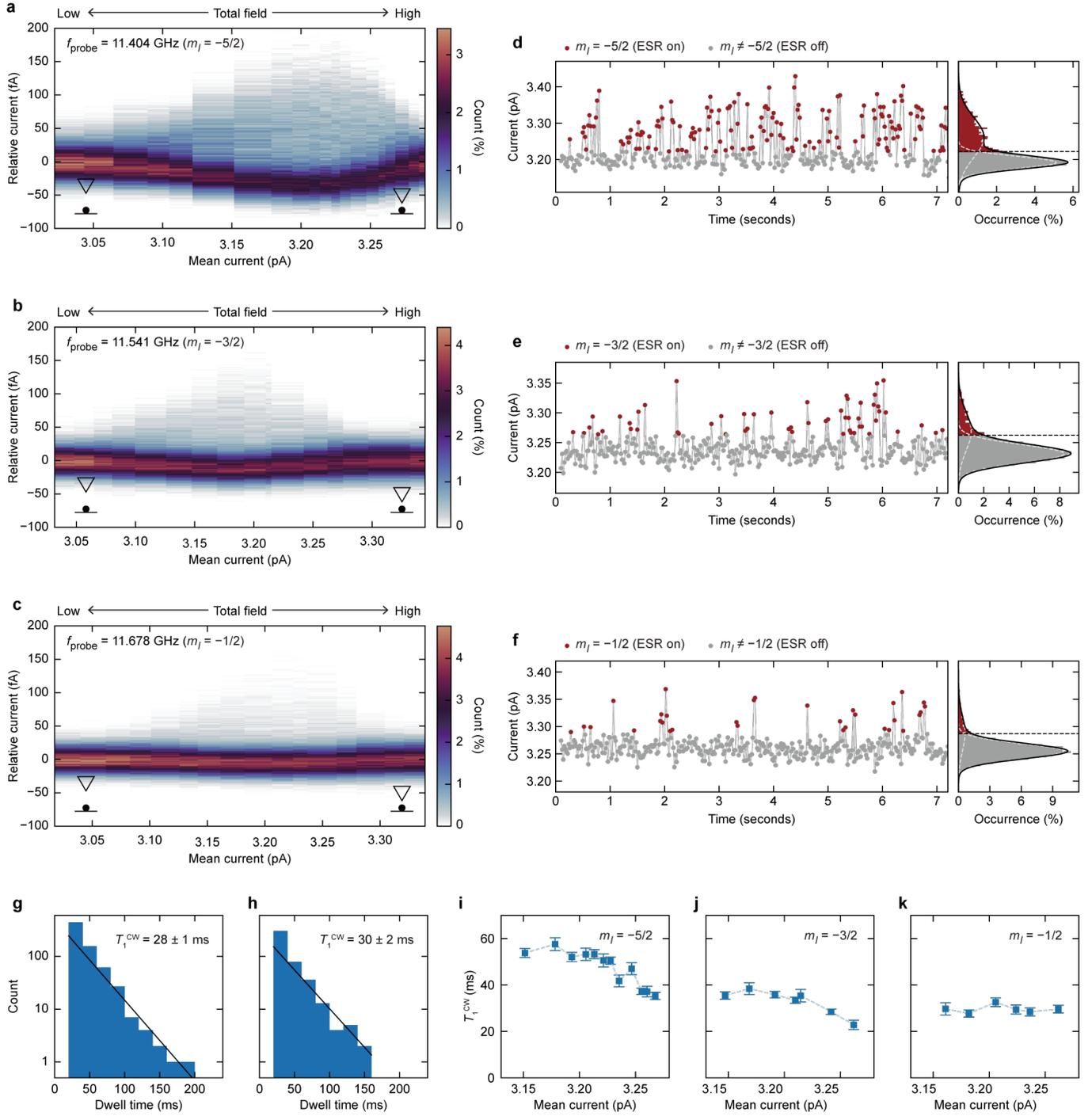

**Supplementary Fig. 2 | Readout of different nuclear spin states at the same tip height. a-c**, Colour maps of current histograms similar to Fig. 2b with different $f_{probe}$ of 11.404, 11.541, and 11.678 GHz, which corresponds to $m_I = -5/2$, $-3/2$, and $-1/2$, respectively. We used the same bias voltage of 60 mV. **d-f**, Representative sections of time traces of the tunnelling current from (a)-(c), respectively. **g** and **h**, Histograms of the distribution of the individual dwell times $t_{dwell}$ of $m_I = -3/2$ and $m_I = -1/2$, respectively. Fitting with an exponential function gives the characteristic dwell time $T_1^{CW}$. **i-k**, $T_1^{CW}$ of $m_I = -5/2$, $-3/2$, and $-1/2$ at different tip heights around the resonance point.

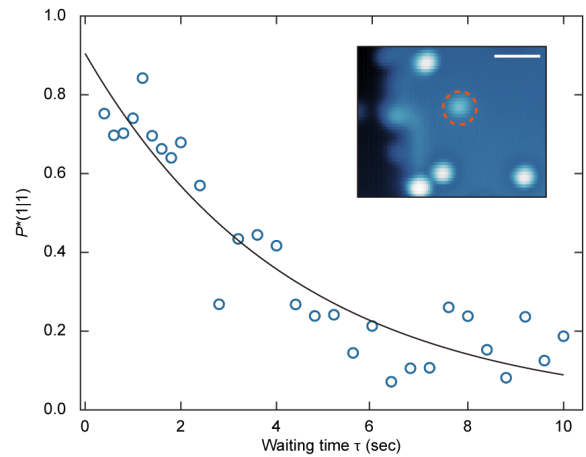

**Supplementary Fig. 3 | Intrinsic nuclear spin lifetime in another $^{49}$Ti atom at the same oxygen binding site.** Corrected conditional probability $P^*(1|1)$ as a function of the waiting time τ. These measurements were performed with a different microtip compared to the tip used for Fig. 4e. With an exponential fit ($Ae^{-\tau/T_1}$, black curve), we measured the intrinsic lifetime $T_1$ of the $m_I = -7/2$ nuclear spin state ($T_1$ = 4.3 ± 0.8 sec, A = 0.90 ± 0.10). Inset: STM topography. The $^{49}$Ti atom is indicated with a dashed circle. Scale bar: 2 nm.

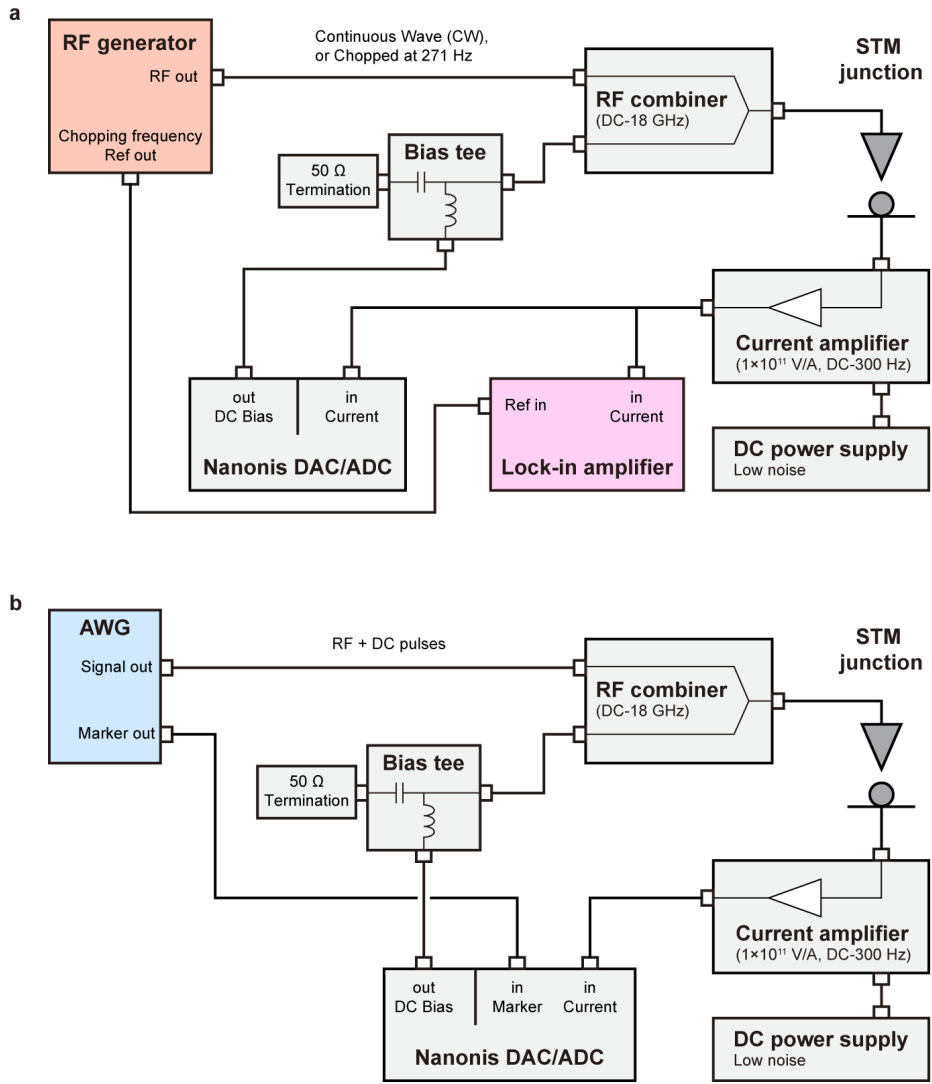

**Supplementary Fig. 4 | Wiring diagrams. a**, Schematic of electronics configuration during CW experiments with current time traces and the conventional ESR-STM frequency sweeps. For the latter, the RF signal is chopped and the current difference between RF on and off is recorded with a lock-in amplifier. **b**, Schematic of electronics configuration during pulse-mode experiments. The RF combiner (Minicircuit ZFRSC-183-S+) with a bandwidth starting at DC is required for the DC sections in the probe pulses at the millisecond scale. We then use our bias tee (Tektronix PSPL5542) as a choke with an inductor to block RF signal from entering the DC bias line, which is not 50-Ohm-matched up to the required RF frequencies. This bias line would otherwise act as a side branch of the main RF circuit and this results in standing waves in the voltage amplitude transfer function. The AC input of the bias tee is terminated with 50 Ohm to prevent standing waves in the short wire branch to the RF combiner.

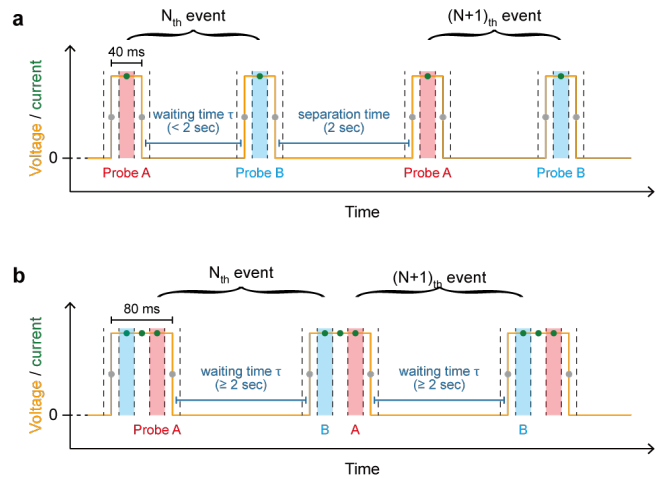

**Supplementary Fig. 5 | Selection of the full-sample probes. a**, Schematic of voltage pulses we send (orange line) and measured data points (dots) for short τ (pulse width = 40 ms). Green and grey circles indicate the full-sample signal and edge signal, respectively. **b**, Similar schematic for long τ (pulse width = 80 ms).

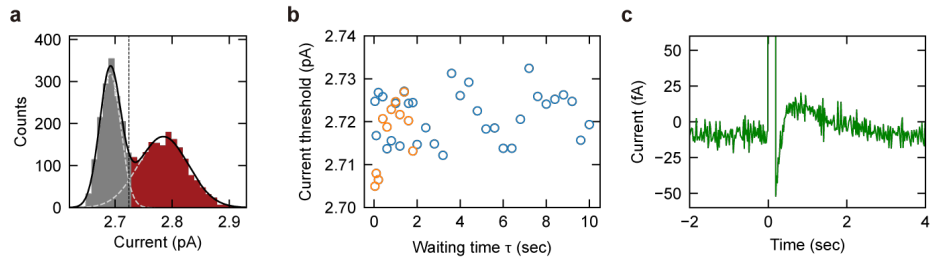

**Supplementary Fig. 6 | Thresholding of probe pulse samples. a**, Current histogram of probe pulse samples for τ = 8 sec fitted with the sum of two Gaussians (black). The intersection point sets the threshold (vertical dashed line) to label probes as either 0 (grey) or 1 (red). **b**, Current threshold determined for different τ, in blue and orange for probe A and probe B samples respectively. Since time traces with τ ≥ 2 sec have pulses operating as both probe A and B, these τ have only one threshold. **c**, Current time trace covering one probe pulse, with the pulse samples cut off vertically to highlight the signal settling after the pulse.

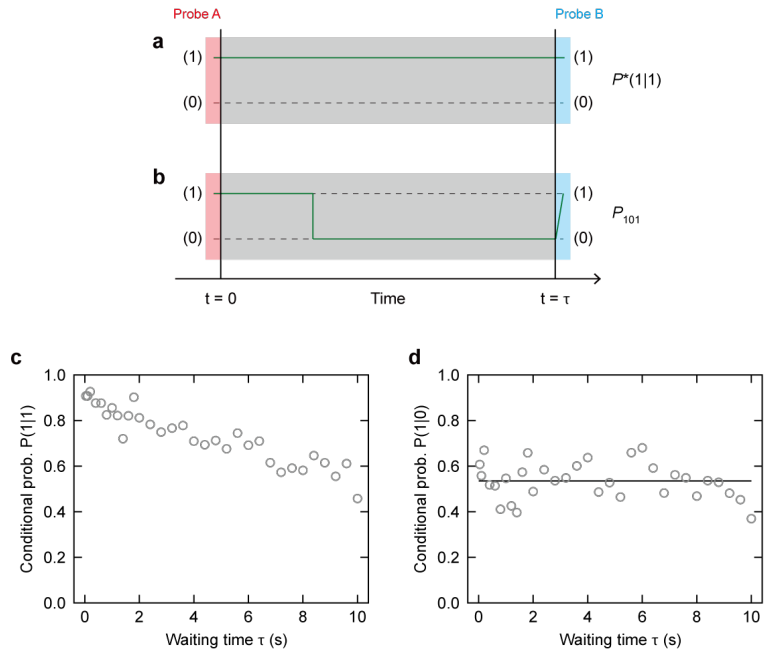

**Supplementary Fig. 7 | Nuclear spin pumping with the second probe. a**, The event for which we want to calculate probability $P^*(1|1)$. **b**, Possible event that the state (1) observed with the probe A decays to the state (0) and is pumped back to the state (1), which corresponds to the $P_{101}$. The green line indicates the nuclear spin state over time. Because we are unaware of what is happening during the waiting time τ (grey box), uncorrected $P(1|1)$ includes the event of (b), that is, $P(1|1) = P^*(1|1) + P_{101}$. **c** and **d**, Conditional probabilities $P(1|1)$ and $P(1|0)$ as a function of waiting time τ, respectively.

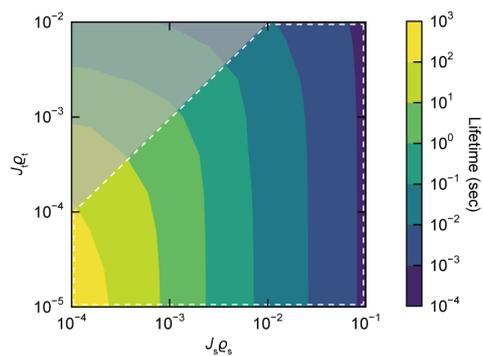

**Supplementary Fig. 8 | Influence of coupling strengths on the nuclear spin lifetime.** Contour plots of lifetimes for the nuclear spin projection $m_I = -7/2$ as a function of $J_s\varrho_s$ and $J_t\varrho_t$. The area of $J_s\varrho_s > J_t\varrho_t$ is highlighted with a dashed box. The transition matrix is calculated under zero bias voltage and no ESR drive.

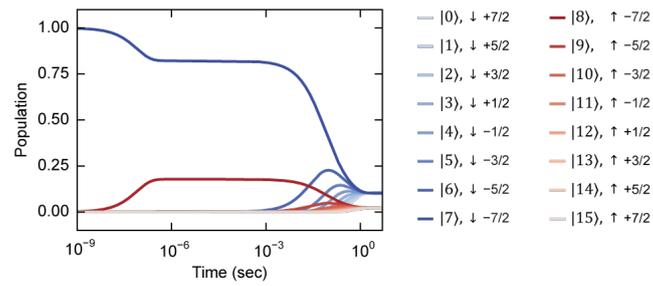

**Supplementary Fig. 9 | Typical simulated time evolution of the eigenstate populations.** Here, $V = 0$ and $\Omega = 0$, thus this simulation represents the undisturbed nuclear spin decay from eigenstate $|7\rangle$. In the legend on the right, eigenstates are labelled by their respective approximate spin product state. Initially, the electron spin evolves to a thermal equilibrium in the $m_I = -7/2$ subspace. On longer timescales, the population is distributed to other nuclear spin projections and the nuclear spin decays.

| $m_s$ | +1/2 ($\|\uparrow\rangle$) | | | | | | | | −1/2 ($\|\downarrow\rangle$) | | | | | | | |
|---|---|---|---|---|---|---|---|---|---|---|---|---|---|---|---|---|
| $m_I$ \ $\|n\rangle$ | +7/2 | +5/2 | +3/2 | +1/2 | −1/2 | −3/2 | −5/2 | −7/2 | −7/2 | −5/2 | −3/2 | −1/2 | +1/2 | +3/2 | +5/2 | +7/2 |
| $\|0\rangle$ | 0.00588 | -0.00098 | | | | | | | | | | | | | -0.00104 | 0.99998 |
| $\|1\rangle$ | 0.00001 | 0.00594 | -0.00129 | | | | | | | | | | | -0.00147 | 0.99998 | 0.00105 |
| $\|2\rangle$ | | -0.00001 | -0.00599 | 0.00146 | | | | | | | | | 0.00176 | -0.99998 | -0.00147 | |
| $\|3\rangle$ | | | 0.00001 | 0.00605 | -0.00153 | | | | | | | -0.00197 | 0.99998 | 0.00177 | | |
| $\|4\rangle$ | | | | -0.00001 | -0.00611 | 0.00149 | | | | | 0.00207 | -0.99998 | -0.00198 | | | |
| $\|5\rangle$ | | | | | 0.00001 | 0.00617 | -0.00135 | | | -0.00203 | 0.99998 | 0.00208 | | | | |
| $\|6\rangle$ | | | | | | -0.00001 | -0.00623 | 0.00104 | 0.00171 | -0.99998 | -0.00204 | | | | | |
| $\|7\rangle$ | | | | | | | 0.00001 | 0.00630 | 0.99998 | 0.00171 | | | | | | |
| $\|8\rangle$ | | | | | | | 0.00102 | 0.99998 | -0.00630 | 0.00103 | | | | | | |
| $\|9\rangle$ | | | | | | -0.00140 | -0.99998 | 0.00101 | -0.00001 | 0.00624 | -0.00133 | | | | | |
| $\|10\rangle$ | | | | | 0.00164 | 0.99998 | -0.00139 | | | 0.00001 | -0.00618 | 0.00147 | | | | |
| $\|11\rangle$ | | | | -0.00178 | -0.99998 | 0.00163 | | | | | -0.00001 | 0.00612 | -0.00150 | | | |
| $\|12\rangle$ | | | -0.00183 | -0.99998 | 0.00177 | | | | | | | -0.00001 | 0.00606 | -0.00144 | | |
| $\|13\rangle$ | | -0.00174 | -0.99998 | 0.00182 | | | | | | | | | -0.00001 | 0.00600 | -0.00128 | |
| $\|14\rangle$ | -0.00142 | -0.99998 | 0.00173 | | | | | | | | | | | -0.00001 | 0.00594 | -0.00096 |
| $\|15\rangle$ | 0.99998 | -0.00141 | | | | | | | | | | | | | 0.00001 | -0.00588 |

**Supplementary Table 1 | The eigenstates of the model Hamiltonian.** The coefficients for each eigenstate are rounded to the fifth decimal place. A grey cell indicates the corresponding coefficient is smaller than 0.00001.